# Normative versus strategic accounts of acknowledgment data: the case of the top-five journals of economics

Alberto Baccini[1*] Eugenio Petrovich[1]

[1]Department of Economics and Statistics, University of Siena, Italy

**Abstract**

Two alternative accounts can be given of the information contained in the acknowledgments of academic publications. According to the mainstream normative account the acknowledgments serve to repay debts towards formal or informal collaborators. According to the strategic account, by contrast, the acknowledgments serve to increase the perceived quality of papers by associating the authors to influential scholars. The two accounts are assessed by analyzing the acknowledgments indexed in Web of Science of 1218 articles published in the "top-five journals" of economics for the years 2015-2019. The analysis is focused on six dimensions: (i) the style of acknowledging texts, (ii) the distribution of mentions, (iii) the identity of the most mentioned acknowledgees, (iv) the shares of highly and lowly mentioned acknowledgees, (v) the hierarchy of the acknowledgment network, and (vi) the correlation at a paper level between intellectual similarity, measured by common references, and social similarity, measured by common acknowledges. Results show that the normative and the strategic account should be considered as valid but partial explanations of acknowledging behavior. Hence, acknowledgments should be used with extreme caution for investigating collaboration practices and they should not be used to produce acknowledgments-based metrics of scholars for evaluative purposes.

**Keywords:** Acknowledgments analysis ● Scientific collaboration ● Quantitative studies of science ● Top five journals of economics ● Directed network analysis ● Intellectual and social similarity ● Symmetric acyclic decomposition

* The research is funded by the Italian Ministry of University, PRIN project: 2017MPXW98. A grant by the Institute For New Economic Thinking, New York Grant ID INO19-00023 is gratefully acknowledged. The funders had no role in study design, data collection and analysis, decision to publish, or preparation of the manuscript.

# 1. Introduction

Collaboration is a distinctive trait of modern science. Scientists collaborate at any scale, from informal lab meetings, to congresses and workshops, up to gigantic clinical trials or massive high-energy physics experiments that involve thousands of scientists and technicians (Brown, 2005; Cronin, 1991; Wray, 2002). Even Nobel prize winners collaborate, frequently more than the average scientist (Zuckerman, 1977). Collaboration has soon attracted the attention of scientometricians, as scientific publications keep various traces of the collaborative process from which they originate. In particular, scientometricians have individuated in the list of authors' names (the by-line of the scientific article) the main textual trace of research collaboration. Co-authorship has thus been elected as the bibliometric proxy of scientific collaboration *par excellence*. Statistics on the incidence of co-authorship have been used for measuring the collaboration rate in the different scientific field and to document how propensity to collaboration changed over time (Cronin, 2001; Larivière et al., 2015; Wuchty et al., 2007), whereas co-authorship networks have become the standard tool to reconstruct patterns of collaboration and to map "invisible collages" (Newman, 2001; Peters & Van Raan, 1991; Zuccala, 2006).

Co-authorship, however, is an imperfect measure of collaboration (Katz & Martin, 1997; Laudel, 2002). On the one hand, it is difficult to assess the contribution of the co-authors, which may vary considerably. In some fields, such a contribution may be limited to providing financial and infrastructural support, without direct involvement in the research process (Larivière et al., 2016). In extreme cases, authorship may be awarded for purely honorary reason ("gift authorship") (Wislar et al., 2011). On the other hand, not all forms of collaborative scientific interaction result in co-authorship. According to Laudel (2002), up to one-third of contributions to research are not rewarded with authorship. Examples of these "invisible contributions" include the work of laboratory technicians and, sometimes, even that of graduate students and statisticians (Cronin et al., 2004; Fogarty, 2020; Shapin, 1989). Furthermore, some collaborative relations that may be warrant authorship in one field may not be sufficient in another, depending on the norms and practices governing authorship attribution (Biagioli & Galison, 2003). Especially in the humanities and the social sciences, where authorship seems to be strictly dependent on the material writing of the publication (Hellqvist, 2009), informal collaboration may be particularly difficult to track by co-authorship, as it rarely results into co-authored publications (Díaz-Faes & Bordons, 2017). In these fields, the assessment of collaboration seems to be highly underestimated when solely co-authorship is considered (Laband & Tollison, 2000; Paul-Hus, Mongeon, et al., 2017). Lastly, in some areas of medical research, the number of co-authors may be constrained by editorial policies of journals that prescribe a maximum number of authors per paper (Alvarez & Caregnato, 2021).

In the light of the shortcomings of co-authorship, a long tradition in library and information science has relied upon the *acknowledgments* of academic publications to obtain a richer picture of collaboration (Desrochers et al., 2017). The acknowledgments are classified as "para-texts" of academic texts, along with titles, headings, illustrations, and dedications (Genette, 1997). They originate from the covering letters accompanying scientific articles, in which scientists thanked patrons and powerful benefactors, and became a standard practice in the 1960s, especially in the Anglo-American academic world (Salager-Meyer et al., 2011). Since these texts frequently mention several *informal collaborators* relevant in the genesis of scientific ideas (Cronin et al., 2003, 2004), they seem a suitable source to complement co-authorship in the study of collaboration practices in science (Paul-Hus, Díaz-Faes, et al., 2017; Paul-Hus, Mongeon, et al., 2017). Sometimes, even formal collaboration may be rewarded by the acknowledgments, as noted by Alvarez and Caregnato (2020).

However, as citations and authorships, also acknowledgements can be given for purely social reasons. An acknowledgement to an important scientist may be less a repayment of an intellectual debt, than a signal to readers and editors that the authors belong to the "right circles" and hence their paper is worth reading and citing (Berg & Faria, 2008). Strategic reasons may guide the decision of mentioning or not mentioning colleagues as they sometimes guide the choice of references. Thus, acknowledgements may suffer from the same limitations related to authorship, with "gift" and "ghost" acknowledgements occurring (Costas & Leeuwen, 2012; Kassirer & Angell, 1991). In the light of this scenario, should the acknowledgments be considered as pure «nano narratives of scholarly collaboration» (Cronin & Franks, 2006, p. 1910) or, alternatively, as social signals of «tribal affiliations and loyalties» (Cronin, 1995, p. 72)?

This interpretative question does not have a purely theoretical interest but addresses the *reliability* of the acknowledgments as sources of data about scientific collaboration. With the increasing availability of acknowledgments in the main citation indexes, advanced large-scale studies of acknowledgments data are becoming possible (see e.g., Giles & Councill, 2004; Paul-Hus, Díaz-Faes, et al., 2017; Paul-Hus, Mongeon, et al., 2017). These studies would benefit from solid methodological and interpretative underpinnings.

This study is an attempt to address this question by a quantitative investigation of the acknowledgments in the field of economics.

The paper is structured as follows. In Section 2 two contrasting interpretations of the acknowledgments' role in the primary communication system of science are delineated. As for citations (Baldi, 1998; Bornmann & Daniel, 2008), they are called the *normative* and the *strategic* accounts of acknowledgments. Section 3 presents the rationale of the quantitative analyses designed for assessing the two interpretations. Sections 4 motivates the choice of economics as case study. Section 5 describes the collection, extraction, and cleaning of data. Results are presented in Section 3 and discussed in Section 4. Lastly, Section 5 concludes.

## 2. Explaining acknowledgments: normative versus strategic accounts

In the literature on the acknowledgments, we suggest distinguishing two main explanatory-interpretative frameworks, the normative and the strategic. These accounts are based on two different ideas of the *function* that the acknowledgments have in the primary communication system of science. In this section, we systematize these two accounts, whose tenets are dispersed in the literature, in order to delineate their "ideal types". We are aware that real studies of acknowledgments frequently blend together elements from both the accounts (see below and the Conclusion section). Yet, a systematization that stresses the differences between the normative and strategic account allows to specify their different predictions and, hence, helps to design the quantitative analyses that will be described in the next section.

According to the normative account, then, the acknowledgments would be used by scientists for recognizing the contribution of colleagues and other informal collaborators (Cronin, 1995; Cronin & Franks, 2006). The normative account assumes that there is a set of tacit norms or common practices that govern the acknowledgment behaviour of authors (Cronin & Overfelt, 1994). These norms prescribe to reward informal contributions by public mentions in academic publications. In this regard, being mentioned in the acknowledgments would be a sign of prestige comparable to being cited and being the author of a scientific publication. Accordingly, Cronin and Weaver-Wozniak (1995) proposed to consider the acknowledgments as a further mechanism by which prestige is distributed in the scientific community. Together, acknowledgments, citations, and authorships would form the «reward triangle of science» (Costas & Leeuwen, 2012; Cronin & Weaver-Wozniak, 1995). Cronin even advanced the idea of creating an "acknowledgments" index that could track the number of acknowledgments each scientist receives (McCain, 2018). According to Oettl (2012), this metric could be used to measure the "helpfulness" to the scientific community, i.e., the propensity to collaborate even without the formal reward of co-authorship. Acknowledgments-based metrics would offer a more balanced assessment of scientific performance than measures focussed only on productivity and citation impact. From the viewpoint of the study of scientific collaboration, the normative account implies that the acknowledgements are a reliable source of data about informal collaboration patterns. The persons mentioned in the acknowledgments can be legitimately considered as sub-authors of the main publication (Patel, 1973) or informal collaborators of the formal authors (Laband & Tollison, 2000), and “gift” and “ghost” acknowledgments are not the norm.

The strategic account of acknowledgments, on the other hand, points out the *instrumental nature* of the acknowledgments and the *strategic motivations* that lie behind the acknowledgment behaviour of researchers. Researchers would not use the acknowledgments to repay informal contributors and reward collegiality. Rather, they would choose the persons to mention based on the effect they may have on editors and readers (Berg & Faria, 2008). In the strategic account, the acknowledgments are interpreted as *signals* used by authors for conveying messages about their affiliation to intellectual schools or academic circles. By trading on the prestige of the persons they mention, authors would increase the perceived quality of their papers, augmenting chances of being published, read, and cited (Coates, 1999). The acknowledgment behaviour would then rest upon strategic considerations rather than credit recognition (Giannoni, 2002). The acknowledgments may be used by authors looking toward the possibility of receiving future favours, as a form of scholarly patronage (Forzetting, 2010). In this setting, the acknowledgments cannot be interpreted straightforwardly as traces of informal collaboration, neither the acknowledgees can be considered as sub-authors by default. According to the strategic account, the number of mentions received by researchers is not a measure of their collegial contribution or helpfulness but of academic power, social influence, or symbolic capital (Brown, 2005). Hence, the strategic account disputes the value of the acknowledgments for the study of scientific collaboration, assuming that “gift” and “ghost” acknowledgments are common phenomena.

In a nutshell, according to the normative account the accumulation of acknowledgements generates the prestige of a scholar; according to the strategic account instead the prestige of a scholar generates the acknowledgements they receive. If it were possible to carry out a simple test of direction of causality, it would also be possible to distinguish which of the two accounts is the correct one. Unfortunately, this simple test is unavailable and the two accounts rest on various empirical observations about acknowledgments practice. However, it should be noted that, in library and information science, the normative account is frequently taken for granted and, thus, is rarely backed up with specific evidence. In fact, it often provides the very rationale for engaging in acknowledgments studies, resulting deeply embedded both in the design of the studies and in the interpretation of the results (see e.g., Paul-Hus, Mongeon, et al., 2017).

Three characteristics of the acknowledgements are generally considered as supporting the normative account. First, acknowledgments are frequently phrased in a thanks-centred vocabulary. "Peer Interactive Communication" (PIC), i.e., the exchange of ideas and comments between authors and acknowledgees, is often explicitly signalled in acknowledgment texts (Cronin, 1995). Costas and Leeuwen (2012) found that, in a sample of around one million of articles from various disciplines featuring an acknowledgments, on average 23% contained PIC-denoting words. In reviews, the percentage increased to 32%. In articles from disciplines such as economics and humanities, the percentage was more than 50%. More recently, Paul-Hus *et al.* (2017) conducted a large-scale study on more than one million acknowledgment texts and found that noun phrases associated with PIC are especially common in the social science and professional fields, whereas in laboratory sciences the role of equipment and materials is more prominent. Alvarez and Caregnato (2020) report that processes of PIC are mentioned in 47.1% of the articles by Brazilian researchers that feature an acknowledgments, with great variances among disciplines: 29% of articles in hematology mention PIC, against 92.2% of the articles in economics. Analogously, Rose (2018) found that 94% of articles in financial economics acknowledge persons as commenters.

Second, journals often provide the authors with detailed guidelines for awarding authorship, which include prescriptions about acknowledgments as well. For instance, the International Committee of Medical Journal Editors (2019) state that contributors that do not meet all the requirements for authorship should be acknowledged explicitly and their contributions specified. Guidelines like these are in fact attempts to institutionalize the norms described by the normative accounts. Hence, they are likely to influence the acknowledgment behaviour of scientists in accordance with the tenets of the normative account.

Third, surveys and interviews of scientists and scholars from different disciplines showed that authors subscribe to the idea of an etiquette governing the writing of acknowledgments (Alvarez & Caregnato, 2020; Cronin & Overfelt, 1994). Even if what constitutes acceptable and unacceptable acknowledgment practice is little codified, Cronin and Overfelt report that authors share an expectation of being acknowledged for contributions to the work of others and likewise expect to give acknowledgments under certain conditions. These expectations attest the presence if not of codified norms, at least of common practices shared in the scientific communities. Correct acknowledgment seems a «question of manners» that is frequently transmitted as tacit knowledge from mentors to students (Cronin & Overfelt, 1994, p. 184). Alvarez and Caregnato (2020) confirmed these results by interviewing twelve researchers from different disciplines affiliated with the Universidade Federal do Rio Grande do Sul in Brazil.

They found that the researchers' acknowledgements behaviour follows ethical and professional, albeit not codified, rules of conduct.

The strategic account, on the other hand, is supported by the following empirical observations.

Berg and Faria (2008) showed that, in economics top journals, there is an inverse correlation between the seniority of the authors and the number of acknowledgees they mention. Senior economists tend to thank less persons than their early-career counterparts. To explain this pattern, Berg and Faria argue that early-career economists use the acknowledgments to signal to editors and readers that they are affiliated with respected scientific circles. By extending the list of acknowledgees, junior authors maximize the chances that readers recognize some of the acknowledgees and hence assess more favourably their paper. Hence, Berg and Faria conclude that the acknowledgments serve to increase the perceived quality of scientific work, rather than to reward informal collaboration. The fact that the vast majority of papers in financial economics acknowledges the editor of the respective journal (Rose, 2018, p. 20) further reinforces the impression that name-recognition and currying favour may be relevant motivations behind concrete acknowledgment behaviour.

Moreover, there is anectodical evidence that, again in the field of economics, the acknowledgments are sometimes used to avoid certain reviewers, on the assumption that editors would not select already acknowledged commenters as referees («Cite you friend, acknowledge your foe»). The fact that guides to academic etiquette in economics explicitly discourage this usage suggests that the behaviour is not so rare in the community (see e.g., Hamermesh, 1992, p. 171).

By the same token, the existence of journal guidelines on the correct attribution of authorships and acknowledgments shows that the concrete practice does not always conform to the principles of the normative account. If the norms would be followed naturally by authors, there would be no need to state them explicitly. Indeed, some suspect cases of inflated acknowledgments are sometimes lamented by editors (e.g., Kassirer & Angell, 1991).

Analogously, Cronin and Overfelt's survey shows that authors are aware both of norms and of the fact that these norms are not always attended: «infractions, great and small, are not uncommon: the existence of a tacit rule set does not ensure that breaches of trust and collegiality will not occur» (Cronin and Overfelt 1984, p.189). Some participants to the survey reported cases in which the acknowledgments were used for strategic positioning (e.g., "Acknowledgement seemed gratuitous; author hoped to enhance reputation by associating with me", p. 173) and cases of "ghost" acknowledgments ("He performed experiments which supported and reinforced the mechanism I had proposed but he gave no acknowledgement when he published his results",

p. 177). Similarly, researchers interviewed by Alvarez and Caregnato (2020) recognize that there are cases in which authors fail to acknowledge the contributions received, even if the lack of formal rules makes impossible to sanction this behaviour.

This array of more and less structured observations about acknowledgment practices show at least that the normative account cannot be taken for granted. A complex bundle of motivations may lie behind the acknowledgments, in addition to reward of collaborators.

Recent research on the acknowledgments, in fact, has recognized the multiple functions of these texts, which stand at the crossroad of both symbolic and collaboration dynamics (Desrochers et al., 2017; see also Paul-Hus & Desrochers, 2019). As to the multiple motivations behind the choice of the persons to mention, for instance, Desrochers and colleagues (2018) note: «Just as the underlying reasons to cite can be diverse, noble, or self-serving, the motivations to acknowledge the support of others can range from flattery and name dropping to the sincere or required demonstration of gratitude upon individuals, organizations, or funding agencies» (p. 233). Large-scale analyses have shown that the acknowledgments cannot be simply considered ad "thank you notes", as they frequently address other topics besides collaboration, such as conflicts of interest, disclaimers, and ethical issues (Paul-Hus, Díaz-Faes, et al., 2017; Paul-Hus & Desrochers, 2019). Funding and funding agencies play also a prominent role and, in the last decade, many studies have investigated the acknowledgments from this angle (see, among others, Alvarez & Caregnato, 2018; Álvarez-Bornstein et al., 2019; Morillo, 2019; Yan et al., 2018; Aagaard et al., 2021; Paul-Hus et al., 2016). The present study fits into this strand of research which aims at shedding light on the multi-dimensionality of the acknowledgments. Specifically, we aim to explore to which extent the normative and the strategic accounts may contribute to explain the concrete acknowledgments behavior of scholars.

# 3. Rationale of the quantitative analyses

Assessing empirically the merits of the two accounts faces challenges analogous to those raised by the corresponding accounts in citation theory. Motivations behind citations and acknowledgments are difficult to investigate, as they are private mental states (Cronin, 1984). Interviews and survey are helpful to access them but are not immune from problems: scientists may be not fully aware of or candid about their motivations, their memory can be at fault, the offered motivations may be post-hoc rationalizations, and so on (Baldi, 1998; Bornmann & Daniel, 2008). Simple quantitative analyses such as counting the number of persons mentioned in the acknowledgments are useful to quantify "acknowledgments intensity" in scientific fields (Cronin et al., 2003) but, alone, do not support either account as they are compatible with both. In fact, the correct interpretation of this kind of statistics is what is at stake.

If the motivations behind the individual acknowledgments are likely to remain unknown, the *aggregate behaviour* generated by a massive amount of acknowledging authors may help to assess the plausibility of the two accounts, as the normative and strategic accounts predict different *population-level features* of the acknowledgments.

In this study, hence, a quantitative approach to the analysis of the acknowledgements is used to assess the plausibility of the two accounts. Specifically, the following aggregate characteristics of the acknowledgments were explored:

1. **Distribution of words in the acknowledgments.** The normative account suggests the prevalence of a thanks-related vocabulary and the presence of PIC-words. However, defenders of the strategic account may argue that words reflect academic etiquette instead of authors' real intentions.
2. **Distribution of mentions to acknowledgees.** The normative account is coherent with two different types of mention distributions. A normal distribution might emerge if the propensity to collaborate is fairly distributed in the population of scholars and hence mentions are not concentrated in few very collaborative individuals. But it is coherent also with a skewed distribution where some scholars have a high propensity to collaborate because of, e.g., their specific skills. In economics, for instance, econometricians might be more likely asked to informally collaborate because of their knowledge of the technical instruments to be applied for analysing data. This might produce preferential attachment mechanisms resulting in a skewed distribution of mentions (Price, 1976; Seglen, 1992). By contrast, the strategic account is coherent only with a skewed distribution of mentions. There are many kinds of scholar that an author would strategically acknowledge such as the editors of a journal for gaining advantages in the referee process, or research group leaders or prominent scholars for signalling his or her own closeness to them. This preferential attachment mechanism may result in a skewed distribution where the most acknowledged scholars have prominent role or academic power or simply higher seniority in the research community considered.
3. **Identification of the most mentioned acknowledgees.** As noted above, the strategic account predicts that scholars with high academic prestige or power should be in the first ranks. A simple test may consist in considering if the most mentioned scholars result to be also editors, i.e., the gatekeepers, of the journals where the articles are published.
4. **Quotas of influential and non-influential acknowledgees.** As Rose and Georg (2021) note, there is little signalling value in mentioning researchers that are relatively unknown. Hence, if most of the acknowledgees that an author mentions receive only few mentions in overall, it is more plausible that they are mentioned to reward their collaboration, as the normative account predicts, rather than for signalling purposes.

5. **Reciprocate acknowledgments and hierarchy in the acknowledgment network.** From a network analysis perspective, the acknowledgment relation is asymmetric: an author acknowledging a scholar is linked to them by an arc pointing from the former to the latter. Nevertheless, the relation between two researchers reciprocating the acknowledgement in their publications is symmetric, as the two authors are linked by two arcs pointing in opposite directions. On the basis of this simple intuition, it is possible to develop a structural analysis of the author-acknowledgees network, hereafter the *acknowledgment network*. First of all: reciprocated acknowledgement may be considered as indicating that two scholars really collaborated and that they are two *peers* (Cronin & Franks, 2006). By contrast, if one acknowledgee receives many mentions from others but never mention back any of the acknowledging colleagues, this may indicate their prestige in the community, rather than their propensity to collaborate. The symmetric and asymmetric relations in the acknowledgement network can be explored by using suitable techniques in view of describing the latent hierarchical structure of the network. The normative account may be considered coherent with a network with a prevalence of reciprocating acknowledgements and a “flat” hierarchy of scholars. The strategic account is instead coherent with a hierarchical structure of the network and with a prevalence of asymmetric acknowledgements pointing toward the highest levels of the hierarchy.
6. **Paper similarity based on acknowledgments vs. paper similarity based on references.** Papers that share several acknowledgees may be intended as similar under the social profile, as they are likely to result from similar academic circles. On the other hand, papers that share many cited references should be similar under the intellectual profile, according to the standard approach of bibliographic coupling (Kessler, 1963; see also Petrovich, 2020). In the normative account, the academic circles of acknowledgees should coincide with intellectual communities of informal collaborators focusing on certain topics. Hence, both the list of acknowledgees and the list of cited references should reflect the same intellectual structure. By contrast, in the strategic account, the social and intellectual profile of a paper should diverge, as the acknowledgees should be chosen based on strategic considerations rather than scientific specialization.

As the discussion of the features shows, the two accounts are frequently compatible with several empirical configurations. Therefore, we do not consider the previous analyses as strict tests of the two accounts. Rather, we are interested in exploring which of the two frameworks appears as more consistent with the data, i.e., which of them provides the "best explanation" (Harman, 1965) for the empirical results.

## 4. Economics as a case-study

The choice of economics as a case-study has a three-fold reason. First, the assessment of the role of the acknowledgees is especially relevant in the study of collaboration in the social sciences and humanities, as in these areas formal co-authorship is less common than in the natural sciences. According to many scholars, acknowledgments permit to capture the amount of collaboration that is lost by co-authorship analysis (Díaz-Faes & Bordons, 2017; Laband & Tollison, 2000; Paul-Hus, Díaz-Faes, et al., 2017; Paul-Hus, Mongeon, et al., 2017).

Second, previous studies found a high acknowledgments intensity in economics. Laband and Tollison (2000) report an average of 15 acknowledgees per paper in top economic journals. More recently, Alvarez and Caregnato (2020) found that 67.4% of articles in economics authored by Brazilian researchers feature an acknowledgment. Interviews to economists confirm that giving and receiving acknowledgements are part of the academic etiquette of the field (Alvarez & Caregnato, 2021). This diffusion of acknowledgment practices warrants that the quantitative analysis does not work on too sparse data.

Lastly, and most importantly, the strategic account was advanced by economists studying acknowledgment behaviour in their own discipline (Berg & Faria, 2008), and, as we saw above, anectodical evidence about strategic usage comes mainly from economics. Moreover, Laband and Tollison (2000) showed that mentioning highly cited and influential economists can have a tangible impact on the career of young economists. They estimated that commentary by a highly accomplished researcher to a novel fellow could be worth a salary increase of almost one thousand dollars for the latter.

Data on the acknowledgments were obtained from the analysis of the articles published between 2015 and 2019 in the so-called Top Five (T5) journals of economics: the *American Economic Review* (AER), *Econometrica* (ECN), the *Journal of Political Economy* (JPE), the *Quarterly Journal of Economics* (QJE), and the *Review of Economic Studies* (RES). According to the description of Heckman and Moktan (2020) they are «"general interest" journals [that] publish papers on a broad range of topics. They are classified in the T5 based on aggregate proxies of journal influence. […] Publication in the T5 journals has become a professional standard. Its pursuit shapes research agendas. For many young economists, if a paper on any topic cannot be published in a T5 outlet, the topic is not worth pursuing. Papers published in non-T5 journals are commonly assumed to have descended into their "mediocre" resting places through a process of trial and failure at the T5s and are discounted accordingly. This mentality is not confined to the young. Habits formed early are hard to break. Pursuit of the T5 has become a way of life for experienced economists as well. Falling out of the T5 is viewed as a sign of professional decline. Decisions about promotions, recognitions, and even salaries are tied to publication counts in the T5». (Heckman & Moktan, 2020, p. 420).

This special role of the T5 suggests that these journals are the best candidates for observing the use of acknowledgement among economists. Given the relevance and diffusion of the T5 in the economic profession, there are good reasons to think that economists are particularly attentive to the use of the acknowledgements in the articles published in the T5. Whatever the reason for thanking someone, the acknowledgements published in the T5 have the largest audience reachable in the economic profession.

As seen above, journals may adopt specific rules for acknowledgements that may have effect on the data under observation. Only two of the T5 journals adopted specific rules. The AER requires that «the main Coeditor in charge will be identified in the acknowledgement note in the published article» (https://www.aeaweb.org/journals/aer/about-aer/editorial-policy). Currently there are 11 coeditors flanking the editor of AER. This may result in a high presence of editors in the list of the acknowledged scholars. The QJE adopt a strict authorship policy, by suggesting that «Any other individuals who made less substantive contributions to the experiment or the writing of the manuscript should be listed in the acknowledgement section» (https://academic.oup.com/qje/pages/Policies). As noted above, this rule represents the codification of the normative account.

## 5. Data collection, extraction, and cleaning

All the research articles published in the T5 were retrieved from Web of Science, for the period 2015-2019.[1] Indeed, acknowledgments data are reliable for publications in the Social Science Citation Index (SSCI) only from 2015 (Grassano et al., 2017). The query resulted in a population of 2,012 articles. However, Web of Science records the acknowledgment only for publications mentioning external funding (Liu et al., 2020; Tang et al., 2017). In our case, they amounted to 1,218 articles, 61% of the total. This limitation is common to all the studies based on WoS data (e.g., Costas & Leeuwen, 2012; Paul-Hus, Díaz-Faes, et al., 2017; Paul-Hus, Mongeon, et al., 2017). This reduction of the set of papers included in the analysis could represent a major shortcoming only if it induced a bias in the final results. This could happen if the set of articles excluded, i.e., the articles that did not declare external funding, used acknowledgements in a structurally different manner from the articles that declared external funding. In the absence of evidence about the existence of this kind of bias, the sample of articles used may be provisionally considered as representative of the way in which acknowledgements are used in T5 journals of economics.

[1] Records were downloaded from Web of Science web interface in May 2020 using the standard University of Siena subscription. Research articles account for 96% of all documents published in the five journals in 2015-2019.

As a first step of analysis, we searched in acknowledgment strings the words indicating review and communication process.[2] In particular, we checked, by using wildcards, the occurrence of the following keywords: "conference", "seminar", "audience", "reviewer", "referee", "editor", and an array of peer interactive communication words.[3] We extracted also the most common lemmas[4] in the whole corpus of acknowledgment texts.

To count acknowledgees mentions and analyse their distribution, the key step was the extraction of the personal names from the text strings. We used for this task the Named Entity Recognition (NER) module of *spaCy* (https://spacy.io/), a Natural Language Processing package for Python. *spaCy* NER recognizes named entities and classifies them in several categories, including persons, organizations, companies, and locations. In this study, we focus mainly on the entities classified as PERSON, i.e., the acknowledgees, as we are interested in the inter-personal dimension of the social context of research. In recent years, however, also funders have attracted the attention of scientometricians as their mention allows to track the research output of different funding strategies, as mentioned above.

A further step of cleaning was needed to correct category misattributions, as *spaCy*'s NER algorithm produces both false positives – entities classified as PERSON that are not persons – and false negatives – entities that are not classified as PERSON but that are persons.

The extracted names, however, could not be used in their raw form because of the presence of several variants of the same name (McCain, 2018). For instance, diminutives were common because of the informal style of several acknowledgments. These variants were merged through a computer-aided cleaning procedure. First, all extracted names were mutually compared using the SequenceMatcher similarity string measure implemented in Python library *Difflib* (https://github.com/python/cpython/blob/main/Lib/difflib.py). All the pairs with a similarity index above 0.8 were retained as possible variants and manually checked. True variants were collapsed to a standard form.

[2] The acknowledgment string was extracted from the FT field of the bibliographic record, which contains the entire text of the acknowledgments.

[3] The list was adapted from (Costas & Leeuwen, 2012) and included: "comment", "suggestion", "communication", "discussion", "reading", "advice", "insight", "inspiration", "inspiring", "correspondence", "feedback", "intellectual debt", "intellectual influence", "conversation", "remark", "discussant", "helpful", "insightful".

[4] A *lemma* is the dictionary form of a word. In English, for example, "reads", "read", and "reading" share "read" as their common lemma. The lemma should not be confused with the *stem*, that is the part of the word that does not change when the word is morphologically inflected.

The last step in the consolidation of the dataset was the removal of those cases where the authors of a paper are mentioned also in the acknowledgments.[5]

The final list of acknowledgees included 7,887 distinct names. The median value of acknowledgees per paper is 12 (mean 13.73, standard deviation = 10.08, skewness = 1.28, excess kurtosis = 2.52). 151 publications do not mention any acknowledgee[6], whereas 25 publications mention more than 40 acknowledgees. The maximum number of acknowledgees mentioned in a single publication is 75. The main descriptive statistics of the dataset are reported in Table 1.

| **Statistics** | | **Value** |
|---|---|---|
| Number of articles in T5 (2015-2019) | | 2,012 |
| Number of articles with acknowledgments (% on all articles) | | 1,218 (61%) |
| Number of distinct authors | | 2,193 |
| Number of articles with acknowledgees (% on articles with acknowledgments) | | 1,067 (88%) |
| Number of distinct acknowledgees | | 7,887 |
| Acknowledgees per paper (only papers with acknowledgees considered) | Median | 12 |
| | Mean | 13.73 |
| | St. Dev. | 10 |
| | Min | 1 |
| | Max | 75 |
| | Skewness | 1.28 |
| | Excess kurtosis | 2.52 |

**Table 1: Descriptive statistics of the dataset.**

Data are available on Zenodo at 10.5281/zenodo.4813214.

# 6. Results

The results of the analyses are organized according to the six dimensions discussed in the Method section.

**Distribution of words in the acknowledgments.** Table 2 shows the most common lemmas found in the entire corpus of acknowledgment texts. Note that using lemmas instead of words allows us to unify variants of the same word (e.g., “comments” and “comment” are unified in the same lemma “comment”).

[5] E.g., the acknowledgments of an article by an article by Ziv Hellman and John Yehuda Levy says: «Ziv Hellman acknowledges research support by Israel Science Foundation Grant 1626/18» (paper n° 194)
[6] In 16 cases, the only "acknowledgees" mentioned were in fact the authors of the papers.

Most common lemmas include words that belong to the semantic spheres of gratitude (“thank”, “helpful”, “grateful”, “acknowledge”) and assistance (“support”, “comment”, “assistance”). Collective occasions of academic discussion such as seminars are also prominent in the corpus (“participant”, “seminar”, “paper”). The role of peer review (“referee”, “anonymous”) and the acknowledgment of financial support (“financial”, “grant”) emerge as well.

| Rank | Lemma | Occurrences | Rank | Lemma | Occurrences |
|---|---|---|---|---|---|
| 1 | University | 2062 | 13 | Economics | 708 |
| 2 | thank | 1648 | 14 | referee | 643 |
| 3 | support | 1428 | 15 | author | 633 |
| 4 | research | 1184 | 16 | helpful | 615 |
| 5 | comment | 1050 | 17 | National | 591 |
| 6 | financial | 1049 | 18 | grateful | 569 |
| 7 | Research | 850 | 19 | Institute | 563 |
| 8 | acknowledge | 818 | 20 | anonymous | 558 |
| 9 | paper | 755 | 21 | School | 552 |
| 10 | participant | 739 | 22 | assistance | 549 |
| 11 | Foundation | 736 | 23 | grant | 531 |
| 12 | seminar | 719 | | | |

**Table 2: Lemmas with more than 500 occurrences in the acknowledgment texts corpus.**

A similar picture is provided by the analysis of the occurrence of specific keywords (Table 3). Economics appears as a community in which intellectual exchange is a common practice.

| Keyword | Articles (abs) | Articles (perc) |
|---|---|---|
| Conference and/or seminar | 763 | 62.6% |
| Audience | 106 | 8.7% |
| Reviewers | 670 | 55.0% |
| Editors | 414 | 34.0% |
| Peer interactive communication words | 983 | 80.7% |

**Table 3: Number and percentage of articles mentioning different types of keywords.**

The acknowledgments of most of the papers (80.7%) include words denoting peer interactive communication. Conferences and seminars are also frequently mentioned, and almost one out of ten papers mention explicitly the audience of seminars or conferences. The reviewers are mentioned in more than half of the papers, the editors in more than one third. In the acknowledgments of 43 articles (3.5% of the total) we find occurrences of all the five families of keywords. Interestingly, only 7 acknowledgments mention negative feedbacks ("critic" or "criticism") and 2 of them define them "constructive criticism". This may indicate that economists overstate positive contributions and downplay conflict in their acknowledgments.[7]

These analyses reveal the prominence of a thanks-related vocabulary in the acknowledgments of economists, a result that is consistent with the normative account. However, they seem to provide a rather idealized representation of academic practices in the field. From this point of view, the defender of the strategic account may argue that, in fact, the acknowledgments are a product of academic etiquette. Moreover, economists may mention conferences and seminars because being accepted at them is a *status symbol* and a signal of competence of the authors, especially when these conferences take place at prestigious universities. Consequently, the contribution of participants and audiences may be overemphasized.

**Distribution of mentions to acknowledgees.** The distribution of mentions among the acknowledgees is positively skewed. The median number of mentions per acknowledgee is 1 (mean = 1.86) and the distribution is characterized by a skewness of 7.87 (excess kurtosis = 108.74). Only 535 acknowledgees receives 5 mentions or more and 5888 (74.6% of the total) are mentioned only once. The maximum number of mentions received by an acknowledgee is 65.

The concentration of mentions in the acknowledgees population is shown by the Lorenz curve in Figure 1. The Lorenz curve plots the cumulative proportion of the mentions ($y$-axis) received by the bottom $x$-percent of the population ($x$-axis). The $x = y$ line represents perfect equality. Figure 1 shows that 75% of the acknowledgees receives only 40% of the mentions.[8]

The Gini coefficient is the ratio of the area comprised between the equality line and the Lorenz curve line over the total area below the equality line. It ranges between 0 and 1, with 0 indicating a perfect equality and 1 the situation of maximal inequality in which all the mentions available were concentrated in one individual. The estimated Gini coefficient is 0.404, indicating inequality in mention distribution among the population of acknowledgees.

---

[7] Similar results were obtained from the analysis of acknowledgments in philosophy, where, however, criticism was mentioned in 7% of the acknowledgments (see Petrovich, forthcoming).

[8] The curve was calculated with R package *ineq* (https://CRAN.R-project.org/package=ineq).

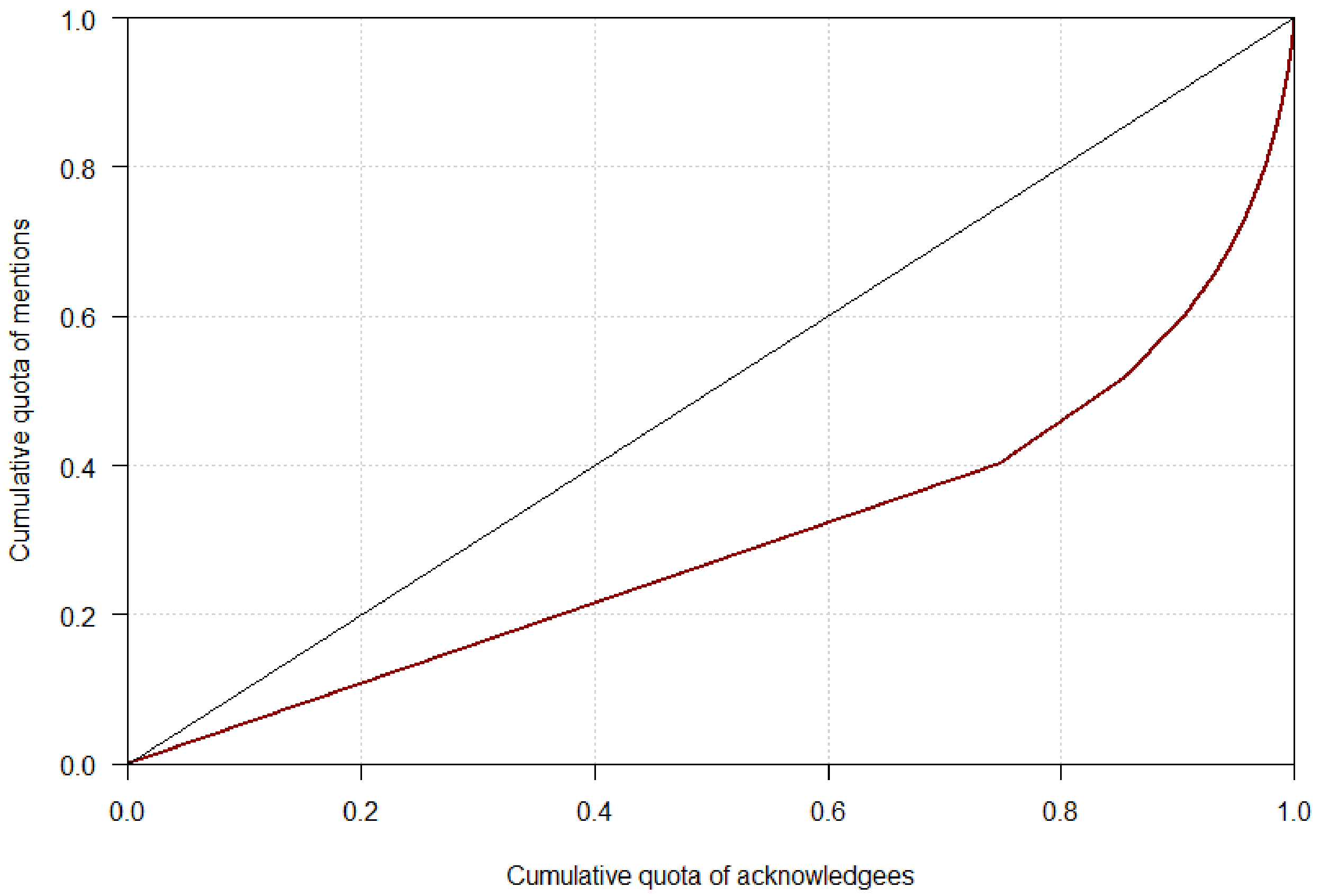

**Figure 1: Lorenz Curve.** *The Lorenz Curve (red line) shows the cumulative proportion of mentions received by the bottom x-percent of the population, against the line of perfect equality (black line).*

If the number of mentions received by an acknowledgee is intended as a measure of their attitude to collaboration, as the normative account suggests, the skewness of the distribution of mentions rule out the hypothesis that this attitude is fairly distributed in the population of acknowledgees. By contrast, a distribution of this kind is most likely explained as resulting from the action of preferential attachments mechanisms (Price, 1976; Seglen, 1992). However, which characteristics of the acknowledgees attract a disproportionate number of mentions remains an open point. In the normative account, these characteristics may be of intellectual type. For instance, econometricians may be preferably chosen for informal collaboration over theoretical economists because of their data analysis skills. Intellectual characteristics of the acknowledgees should then account for the concentration of mentions in relatively few scholars. In the strategic account, by contrast, it is the signaling value of acknowledgees that determine their mention. This value depends on academic prominence, symbolic power, and other non-intellectual features.

These considerations suggest looking at the right tail of the distribution, where the most mentioned acknowledgees are placed, to better understand whether their profile corresponds to that suggested by the strategic account.

**Identification of the most mentioned acknowledgees.** Table 4 shows the list of the most acknowledged economists, i.e., economists with more than 20 mentions ($n$ = 21). They are all male and 18 of them (86%) are or have been members of a T5 journal editorial board, 7 even belong to multiple boards. Most of them are the recipient of several prestigious awards in economics and two (Heckman and Banerjee) received the Sveriges Riksbank Prize in Economic Sciences in Memory of Alfred Nobel. They obtained heir Ph.D. at highly prestigious universities in the United States, where they now hold distinguished professorships (mainly at Harvard and MIT).

| Rank | Researcher | Mentions | Journal editorial board* | Prizes | Alma mater (PhD) |
|---|---|---|---|---|---|
| 1 | Katz, Lawrence F. | 65 | QJE | | Massachusetts Institute of Technology (USA) |
| 2 | Shleifer, Andrei | 60 | QJE | John Bates Clark Medal (1999) | Massachusetts Institute of Technology (USA) |
| 3 | Shapiro, Jesse M. | 47 | QJE; AER | | Harvard University (USA) |
| 4 | Acemoglu, Daron | 42 | ECN** | John Bates Clark Medal (2005)<br>John von Neumann Award (2007)<br>Erwin Plein Nemmers Prize in Economics (2012) | London School of Economics (UK) |
| 5 | Chetty, Raj | 39 | - | John Bates Clark Medal (2013)<br>Infosys Prize (2020) | Harvard University (USA) |
| 6 | DellaVigna, Stefano | 34 | AER | | Harvard University (USA) |
| 7 | Heckman, James J. | 33 | JPE | John Bates Clark Medal (1983)<br>Sveriges Riksbank Prize in Economic Sciences in Memory of Alfred Nobel (2000)<br>Frisch Medal (2014) | Princeton University (USA) |
| 8 | Samuelson, Larry | 28 | AER; ECN** | | University of Illinois (USA) |
| 9 | Kline, Patrick | 27 | AER; ECN; RES | IZA Young Labor Economist Award (2020) | University of Michigan at Ann Arbor (USA) |
| 9 | Finkelstein, Amy | 27 | AER | John Bates Clark Medal (2012)<br>Elaine Bennett Research Prize (2008) | Massachusetts Institute of Technology (USA) |
| 9 | Card, David | 27 | QJE | John Bates Clark Medal (1995)<br>Frisch Medal (2008) | Princeton University (USA) |
| 10 | Gentzkow, Matthew | 25 | QJE | John Bates Clark Medal (2014) | Harvard University (USA) |
| 10 | Glaeser, Edward L. | 25 | QJE** | | University of Chicago (USA) |
| 10 | Horner, Johannes | 25 | AER | | University of Pennsylvania (USA) |
| 11 | Gabaix, Xavier | 24 | ECN; QJE | Bernacer Prize (2010)<br>Fischer Black Prize (2011)<br>Lagrange Prize (2012) | Harvard University (USA) |
| 12 | Bonhomme, Stephane | 23 | RES** | | University of Paris I, Panthéon-Sorbonne (France) |
| 13 | Notowidigdo, Matthew J. | 22 | QJE | | Massachusetts Institute of Technology (USA) |
| 13 | Fudenberg, Drew | 22 | ECN**; QJE**; RES** | | Massachusetts Institute of Technology (USA) |
| 13 | Bloom, Nicholas | 22 | ECN; QJE | Frisch Medal (2010) Germán Bernácer Prize (2012) | University College London (UK) |
| 14 | Laibson, David | 21 | - | | Massachusetts Institute of Technology (USA) |
| 14 | Banerjee, Abhijit | 21 | AER**; QJE**; RES** | Sveriges Riksbank Prize in Economic Sciences in Memory of Alfred Nobel (2019) | Harvard University (USA) |

**Table 4. Ranking of top acknowledgees**. *The table shows the acknowledgees with more than 20 mentions in the corpus (*n *= 21), along with the journal boards they are member of, the prizes they received, and the institution where they obtained the Ph.D. *In 2019 **In the years before 2015-2019.*

The prominent place in the ranking of economists that play the role of gate-keepers of the field is consistent with the predictions of the strategic account.

**Quotas of influential and non-influential acknowledgees.** In the same acknowledgments, however, acknowledgees with different visibility may be mentioned together. As Rose and Georg (2021) note, there is little signalling value in mentioning persons that are little visible in the community. Hence, if most of the acknowledgees in the individual acknowledgments are in fact lowly mentioned in the overall corpus, this may be a sign that authors do not choose their acknowledgees based on visibility, influence, or other forms of symbolic power.

The average number of highly visible acknowledgees (i.e., the acknowledgments with 10 mentions or more) per paper is 3.5, whereas the average number of less visible acknowledgees (i.e., the acknowledgees with less than 10 mentions) is 11.3. The average quota of highly mentioned acknowledgees per paper is 18%, i.e., less than 1 out of 5 acknowledgees in the average acknowledgment belong to the group of highly mentioned. 11 papers mention only highly mentioned acknowledgees, but for 7 of them, they are the only mentioned acknowledgee. By contrast, 293 papers do not mention any highly mentioned acknowledgees. These papers have an average of 6.7 acknowledgees per acknowledgments. The average quota of 1-mention acknowledgees per paper is 42.7%, i.e., almost the half of the acknowledgees mentioned in the average paper do not appear in any other acknowledgments of the corpus. 87 papers mention only 1-mention acknowledgees. These results are in line with Rose and Georg's criticism of the strategic account, that would fail to explain why most of the acknowledgees are in fact non-influential persons. The presence of these lowly mentioned acknowledgees is easily explained by the normative account, as the most likely explanation of their mention is that authors want to reward publicly their collaboration rather than send signals. Yet, by adopting the strategic account, it may be argued that prestige sometimes has a local dimension. A researcher that receives a lower number of mentions in the overall corpus may be nonetheless prominent in a subfield or in a local setting (e.g., a department or a school of thought). Hence, even the mentions to lowly mentioned economists may be signals, that are sent to local communities instead of the discipline at large.

**Reciprocate acknowledgments and hierarchy in the acknowledgment network.** As said in Section 3, the acknowledgment network can be modelled as directed graph in which nodes represent researchers giving and receiving acknowledgments and arc the asymmetric relation of mentioning, whose direction point from the acknowledging author to the mentioned acknowledgee. The in-degree of a node is the number of mentions they receive; the out-degree is the number of acknowledgements they send. A node with indegree equal to zero is a scholar appearing only in the set of the authors of papers; a node with outdegree equal to zero is a scholar appearing only in the set of acknowledgees. To account for multi-authored papers, the weight of an arc is fractioned according to the number of co-authors of the acknowledging paper, so that, for instance, the authors of a paper with 4 co-authors, will be connected to the acknowledgees mentioned in their paper each one with a weight of 1/4 (cf. Rose & Georg, 2021; and Khabsa et al., 2012), and obviously each acknowledge receives 1 mention.

The representation of the network as a directed graph permits to deploy some techniques for analysing its hierarchical structure, if any. The classical instrument for analysing the hierarchical structure of a directed network consists in the study of triads (Wasserman & Faust, 1994), where a triad is composed by three nodes and their links. The triad census consists in counting the possible 16 types of triads in a network and in comparing this distribution of triads with the expected distribution for a random network of the same size. The prevalence of one or more types of triads permits to infer the macro-structure of the network, according to a suitable classification (De Nooy et al., 2018; Johnsen, 1985). In our case, the triadic census, reported in Appendix, does not allow a univocal classification of the acknowledgement network.

Therefore, the technique of *symmetric-acyclic decomposition* is applied for analysing the hierarchical structure of the network. The technique was developed by Doreian and colleagues (2000) and it is applied by using the algorithm available in *Pajek* (De Nooy et al., 2018). The basic idea is that nodes that are linked by symmetric bi-directional arcs directly or indirectly belong to one cluster and to the same rank. In the considered network, this means that when two authors acknowledge each other, i.e., they are connected by a couple of arcs pointing in opposite directions, they are attributed to the same cluster. By contrast, when an author acknowledges a researcher, who does not reciprocate the acknowledgment, the two nodes are not considered as belonging to the same cluster. After having individuated clusters of nodes, they are ranked by considering the direction of arcs linking them.

The application of the symmetric-acyclic decomposition yields a hierarchy composed by 8 partially ranked clusters of nodes as represented in Figure 2. Clusters are identified by different colours and are organized in layers from left to right; hereafter each cluster is identified by its colour or by a letter from A to H.

The first cluster individuated by the symmetric-acyclic decomposition algorithm is the purple one (E) with 850 nodes. It is composed by two types of nodes: 369 nodes directly linked by symmetric bi-directional arcs and forming 47 clusters of mutually acknowledging scholars. For finding the second type of nodes, the algorithm replaces each of the 47 clusters with a shrunk node inheriting all the arcs incident with the original nodes. The second type of nodes in the purple cluster is composed by 481 nodes symmetrically linked to at least one of the 47 shrunk nodes. Namely, a node of this second type is linked by a bi-directional arc to a shrunk node: this node receives an acknowledgement by a member of one of the 47 clusters and it acknowledges another member of the same cluster.

The purple cluster contains all the symmetric arcs of the acknowledgement network. Thus, all the asymmetric, i.e., not reciprocated, acknowledgements are situated between the remaining clusters. In particular, the cluster are ordinated so that all arcs point from left to right. Consider for example the red cluster A: it contains 1,108 nodes. Each node represents a scholar never acknowledged by others and acknowledging one or more scholars in a cluster located further to the right. The green cluster B individuates scholars who are acknowledged by members of the red cluster only or who acknowledge scholars in clusters located further to the right, and so on until purple cluster. The scholars of the purple cluster are acknowledged by members of the clusters to their left; they mutually acknowledge other members of the purple cluster and, finally, they mention scholars of the clusters to their right.

The size of a node in Panel A of Figure 2 is proportional to the weighted indegree of a node, i.e., the weighted sum of acknowledgement received, whereas the size of nodes in Panel B is proportional to the weighted numbers of acknowledgements generated by a scholar. The visual comparison of the two panels shows that the big part of scholars generating acknowledgements belong to clusters A (red) and E (purple). The clusters C, D, F, G and H, on the other hand, are composed by a strong majority of acknowledged scholars never acknowledging others.

A global view of the structure of the acknowledgement network is drawn in Figure 3. Here each node represents a cluster. The light blue cluster F collects the big part of the acknowledgements of the network. The purple cluster E generates the major flow of asymmetric acknowledgement toward members of F cluster. Recall that cluster E is the only one that generates mutual acknowledgements among its members. From its part, the F cluster generates few acknowledgements.

In network analysis jargon, the central purple cluster E is a strong component of the acknowledgement network, i.e. a maximal strongly connected subnetwork where there is a directed path in both directions between every pair of nodes (Newman, 2018, p. 135). The last step of the analysis consists in verifying if this strong component is in turn a symmetric cluster. In this case the strong component is asymmetric, as the purple cluster contains in turn a hierarchy of nodes. Figure 4 shows the subset of 240 nodes linked directly or indirectly by symmetric acknowledgments. Note that all the scholars represented in Figure 4 are authors of at least a paper published in one of the Top5 journals. Note also that the cluster contains all the most acknowledged scholars listed in Table 4.

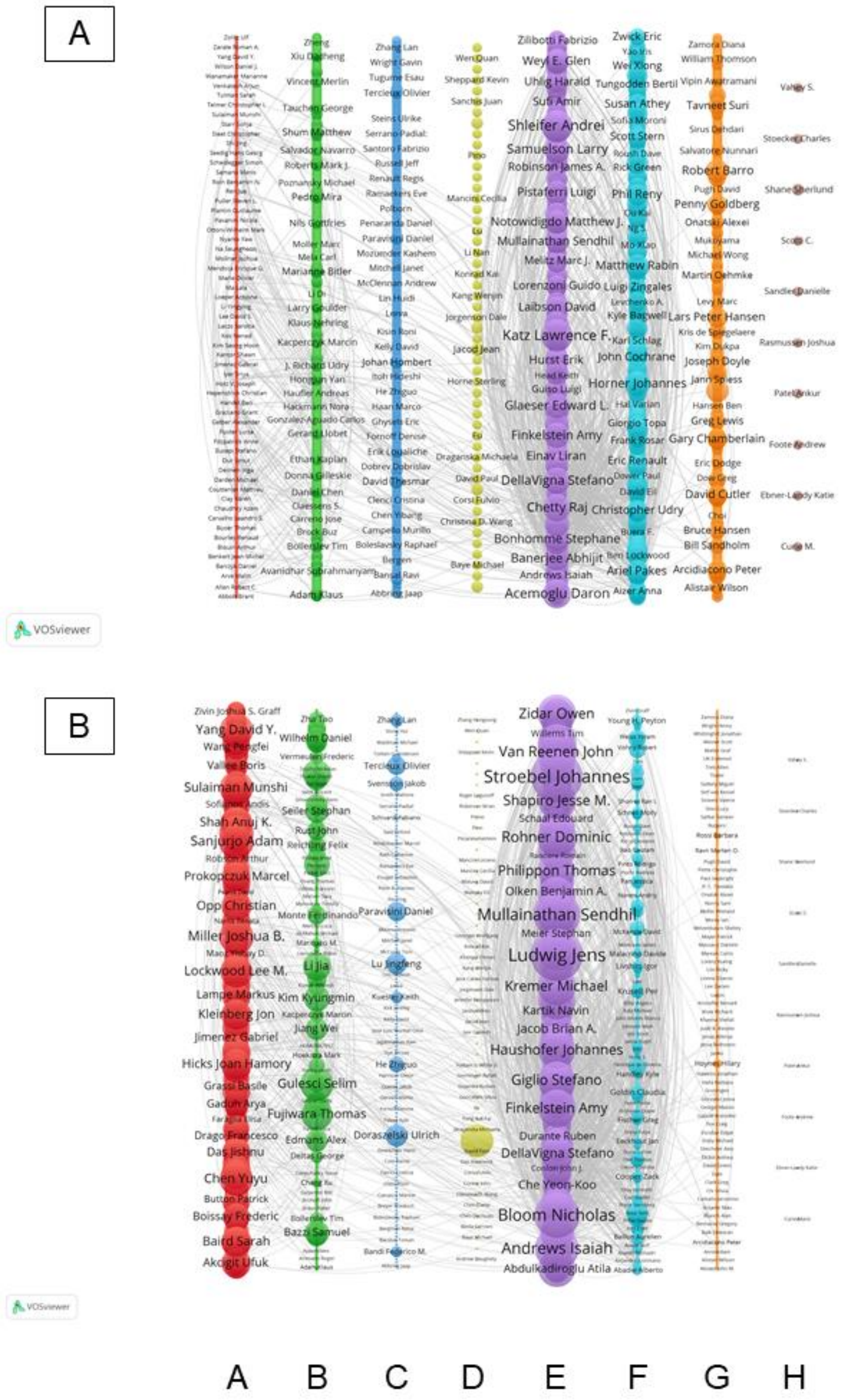

**Figure 2. The symmetric-acyclic decomposition of the acknowledgement network for the top-five journals of economics**. *Clusters of scholars are organized in layers from left to right and identified by different colours. In Panel A, the dimension of a node and of its label are proportional to its weighted indegree, i.e., to the weighted number of acknowledgements received by a scholar. In Panel B the dimension of a node and its label are proportional to its outdegree, i.e., to the weighted number of acknowledgements a scholar has made in their works. In both panel links between cluster point from left to right, i.e., indicate acknowledgement from scholars on a cluster on the right to scholars on the left. The only exception is the purple clusters originating also symmetric links, i.e., acknowledgements toward other members of the purple clusters. Graph are drawn by using Pajek. VOSviewer* (van Eck & Waltman, 2010) *is used for the final layout. Only about a thousand of links are visualized to improve readability.*

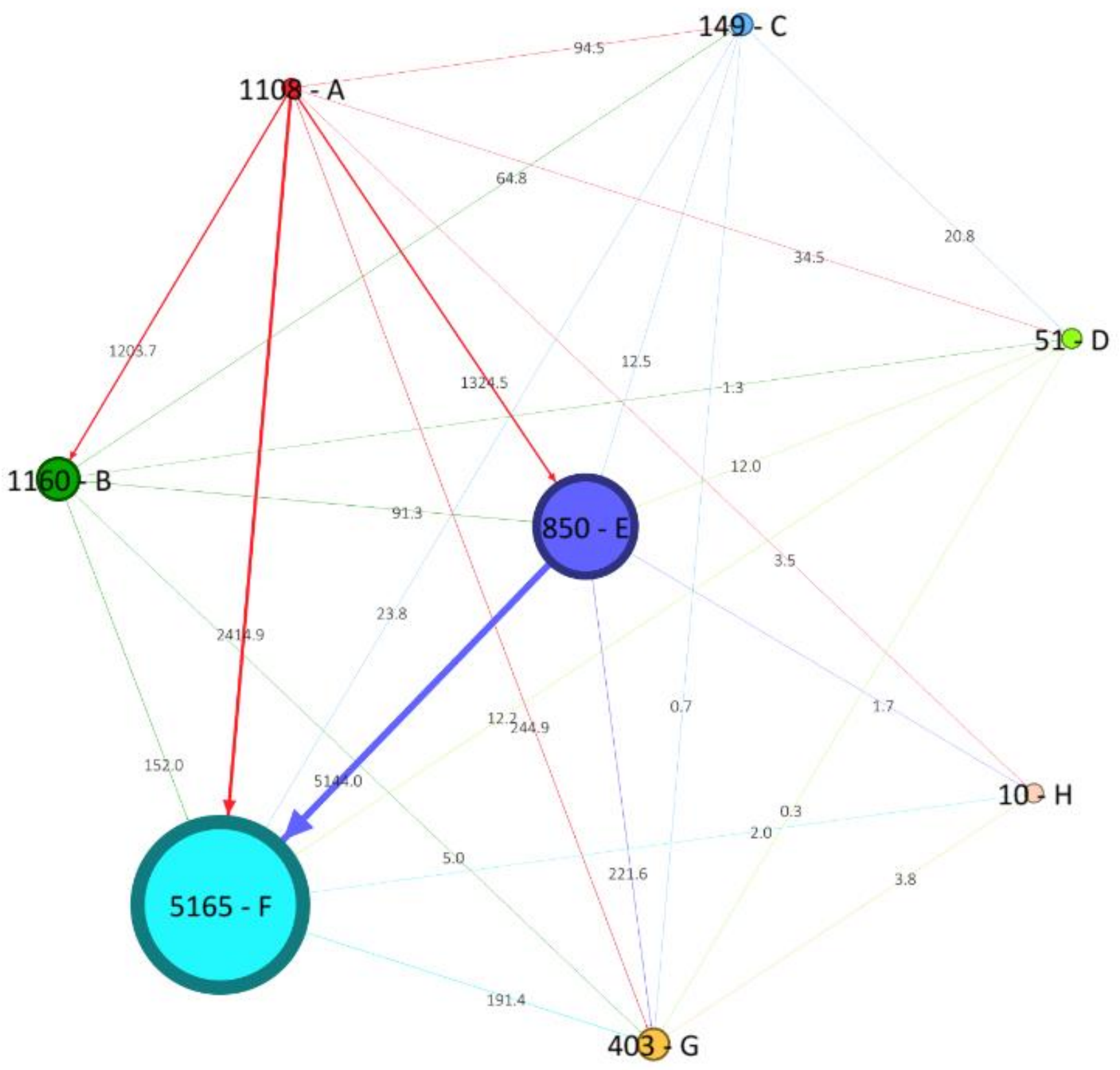

**Figure 3. A global view of the symmetric-acyclic decomposition**. *Each node is a cluster coloured as in Figure 2, the uppercase letter indicates the cluster, and the number corresponds to the number of scholars belonging to the cluster. Size of the nodes is proportional to weighted in-degree of the cluster, i.e., the weighted number of acknowledgements received by the scholars of the cluster. The arcs indicate the direction of the acknowledgements: the arcs are coloured according to the colour of the acknowledging clusters. The number near the arcs indicates the weighted number of acknowledgements generated from the scholars of a cluster toward the other cluster. For example, the red arcs pointing from cluster A to B indicates that scholars in clusters A mention 1230.7 times the scholars in cluster B. The figure is manually adapted starting from Fruchterman-Reingold algorithm by Gephi* (Bastian et al., 2009).

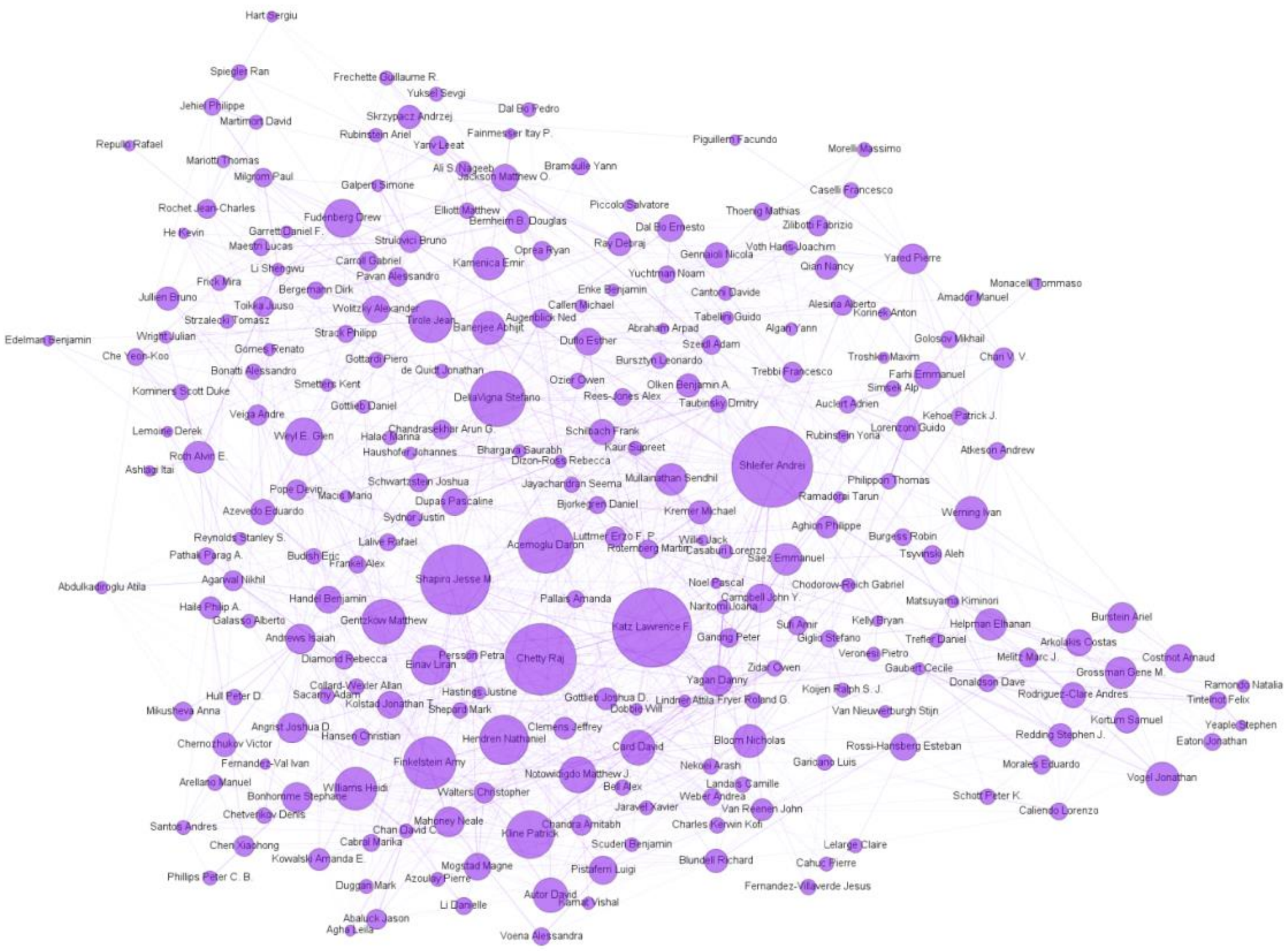

**Figure 4. The cluster of scholars linked by symmetric acknowledgement relation**. *The size of a node is proportional to its weighted indegree in the whole network, i.e., the (weighted) number of acknowledgements received. The figure is energized by ForceAtlas 2 algorithm by Gephi software.*

The symmetric-acyclic decomposition of the acknowledgement network shows a stratified classification of scholars. On the left of the purple clusters there are 4 clusters of scholars (2,468 (27.8%) out of 8,896) never acknowledged by others or acknowledged by not-acknowledged scholars only. All the relations established by the acknowledgements are asymmetric, by indicating a dominance of scholars on the purple cluster over scholars on the left. It appears rather implausible that all these acknowledgements indicate sub-authorship relations, since it appears very implausible that scholars of the purple cluster, accustomed to reciprocating acknowledgements, never reciprocate authors on the left. Hence, it may be conjectured that a big part of the arcs pointing from clusters A, B, C, D to clusters E, F, G, H is strategical. Nevertheless, the scholars in the clusters A-D determine with their acknowledgment choices the centrality of scholars of the purple cluster E in the network.

In turn, the scholars in the purple cluster acknowledge scholars in the clusters to their right, who do not reciprocate. It can be conjectured that these asymmetric acknowledgements may originate not only by collaboration relations. They can originate as tribute of deference toward prestigious scholars who do not reciprocate, for instance simply because they ceased their publication activity. They can originate also as signals indicating that a young scholar is a candidate to be co-opted inside the purple circle of symmetric acknowledgements.

The purple cluster is in turn hierarchically ordered. The most mentioned scholars of the whole network are inside the purple clusters and many of them are also editors of the top five journal of economics.

**Paper similarity based on acknowledgments vs. paper similarity based on references.** Scienometricians have developed many measures of "paper similarity" (Todeschini & Baccini, 2016). The classic bibliographic coupling strength is considered as a measure of intellectual similarity. The intellectual similarity between a pair of paper is indicated as $I(a, b)$ and it is measured by using the Jaccard similarity applied to the sets of references of the two papers $R_a$ and $R_a$, as:

$$I(a,b) = \frac{|R_a \cap R_b|}{|R_a \cup R_b|}$$

Where $|\cdot|$ denotes the cardinality of a set.

Analogously, similarity between two papers can be computed by considering the sets of the acknowledgees they mention. This similarity can be interpreted as a measure of social proximity of the two papers. It is defined as:

$$S(a,b) = \frac{|K_a \cap K_b|}{|K_a \cup K_b|}$$

Where $|\cdot|$ denotes the cardinality of the set, $K_a$ is the set of acknowledgees mentioned by paper $a$ and $K_b$ is the set of acknowledgees mentioned by paper $b$.

Both similarities are defined between a minimum value of 0 and a maximum of 1. The minimum value indicates that the two papers are completely dissimilar since they have no common references or common acknowledgees; the maximum value indicates that the two papers are the same paper, since they have an identical set of references or an identical set of acknowledgees.

Consider now the set of papers $P$ that cite at least one reference and mention at least one acknowledgee. In our corpus, there are 1067 papers satisfying this condition. They are used as nodes in two distinct networks. In the first network $G_s$, two nodes are connected if they share at least one acknowledgee, and the weight of the link equals the social similarity $S(a, b)$ between the two papers. Let us call $G_s$ the *acknowledgment coupling network* of papers $P$. In the second network, $G_i$, the nodes are connected if they share at least one reference, and the weight of the link equals the intellectual similarity $I(a, b)$ between the two papers. $G_i$ is the well-known *bibliographic coupling network* of papers $P$. Note that these two networks may also be intended as the two layers of a multi-plex network insisting on $P$ nodes, the former based on the social similarity between papers, the latter based on the intellectual similarity between them.

If the normative account is correct, we expect that the two networks are characterized by a similar structure. In fact, if the ties between authors and acknowledgees are of intellectual type, as the normative account assumes, we expect that authors ask the advice of colleagues competent on the topic they address in their papers. Accordingly, two papers that share several acknowledgees should address similar topics. By contrast, if the acknowledgees are mentioned because of their signaling value, independently of their intellectual specialization, the bibliographic and social profiles of papers should be unrelated.

Note that $G_s$ has an average link weight of 0.0369 (st. dev. = 0.0239, min = 0.008, max = 0.700), that is two times the average link weight of $G_i$, that equals 0.0169 (st. dev. = 0.0164, min = 0.002, max = 0.381), showing that papers tend to be, on average, more similar under the social than intellectual profile. The information about the similarities of the papers in the two networks can be organized in two square symmetric similarity matrices, one for each network. It is therefore possible to compare the two similarity matrices by using the generalized distance correlation $R_d$ suggested by (Székely et al., 2007) to assess the correlation between the similarity matrices of the two networks. This measure is interpreted similarly to the squared Pearson correlation coefficient. Hence, its square root $\sqrt{R_d}$, which ranges between 0 and 1, may be seen as a generalization of the usual correlation coefficient. Values closer to 1 indicate a strong association, whereas values closer to 0 no or weak association. In our case, we obtained an intermediate value of 0.4929, which suggest a medium correlation between the similarity matrices.[9]

A further way to assess the similarity between the two networks is to compare their community structure. If the communities found in $G_s$ are the same as the communities found in $G_i$, we can say that the two measures of paper similarity essentially give rise to the same partition of papers $P$. Such a situation would support the normative account. By contrast, if the community structures diverge, the defender of the strategic account may argue that non-intellectual factors intervene in the choice of the acknowledgees. Specifically, acknowledgees may be chosen because of their prestige in the community rather than their intellectual closeness with the authors.

[9] The generalized distance correlation was calculated with R package *energy* *https://github.com/mariarizzo/energy* ).

The classic Louvain algorithm as implemented in Pajek individuates 97 communities in $G_s$. Most of them, however, comprise only 1 node. The six main communities include 969 nodes (91% of the total). In $G_i$, the algorithm finds 18 communities. Again, the seven main communities include 1056 nodes (99% of the total). In the following, we focus only on the six main communities based on social similarity and on the seven main communities based on intellectual similarities. They can be considered as alternative partitions or classifications of the same set of papers.

The Chi-squared test rejects the null hypothesis that the two classifications are independent ($\chi^2(30, N = 964) = 1146.883,\ p < 0.000$) and the Cramer's V yields an effect size of 0.45, which attests a medium association strength between the two.

<table>
<tr><th colspan="2" rowspan="2"></th><th colspan="7">Social communities</th><th rowspan="2">Total</th></tr>
<tr><th>1</th><th>2</th><th>3</th><th>4</th><th>5</th><th>6</th><th>7</th></tr>
<tr><th rowspan="6">Intellectual communities</th><th>A</th><td>75</td><td>6</td><td>12</td><td>4</td><td>6</td><td>23</td><td>27</td><td>153</td></tr>
<tr><th>B</th><td>1</td><td>20</td><td>10</td><td>17</td><td>121</td><td>14</td><td>25</td><td>208</td></tr>
<tr><th>C</th><td>8</td><td>15</td><td>28</td><td>2</td><td>4</td><td>16</td><td>21</td><td>94</td></tr>
<tr><th>D</th><td>14</td><td>25</td><td>33</td><td>8</td><td>8</td><td>96</td><td>19</td><td>203</td></tr>
<tr><th>E</th><td>22</td><td>7</td><td>54</td><td>103</td><td>8</td><td>22</td><td>7</td><td>223</td></tr>
<tr><th>F</th><td>2</td><td>58</td><td>4</td><td>3</td><td>5</td><td>8</td><td>3</td><td>83</td></tr>
<tr><th colspan="2">Total</th><td>122</td><td>131</td><td>141</td><td>137</td><td>152</td><td>179</td><td>102</td><td>964</td></tr>
</table>

**Table 5: Contingency table between intellectual communities (columns) and social communities (rows).** *Cells with the highest frequency in the row are highlighted in green.*

To better understand the relationship between individual intellectual and social communities, the contingency table showing crossed frequencies between the two classifications is reported in Table 5. Communities based on intellectual similarity are on the columns, communities based on social similarity are on the rows.

The observed frequencies show that papers are rather dispersed among the different communities. Some social communities, however, seem more associated with specific intellectual communities. For instance, 121 out of 208 papers (58%) belonging to social community B appear in intellectual community 5.

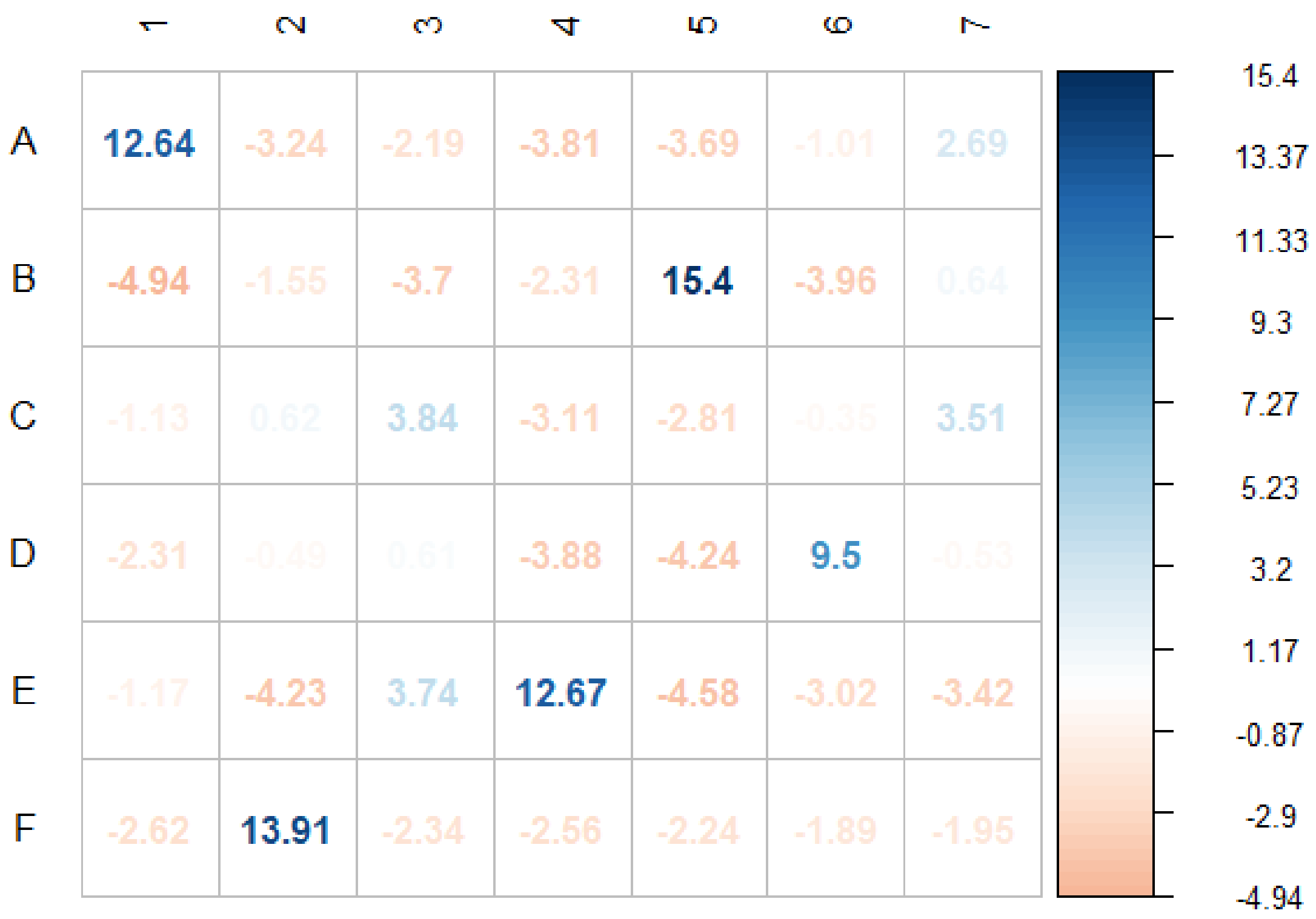

**Figure 5: Association plot between intellectual communities (columns) and social communities (rows).** *The association plot shows the Pearson residuals associated with each pair of communities. Positive residuals are in blue, negative residuals in red. Residuals close to zero are in white. The plot was produced with R package* corrplot(*https://github.com/taiyun/corrplot*).

To quantify the association strength between each pair of communities, the association plot in Figure 5 reports the Pearson residuals associated with each pair.[10] Positive residuals indicate that the observed frequency is higher than expected, and hence that a positive correlation exists between the two communities. On the other hand, negative residuals indicate that the observed frequency is lower than expected and, hence, that the two communities are negatively correlated (Agresti, 2007).

Figure 5 confirms that, even if some social communities have their intellectual counterpart, two of them (C and D) are weakly associated with any specific intellectual community. The

[10] Pearson residuals are defined as:

$$r = \frac{f_o - f_e}{\sqrt{f_e}}$$

Where $f_o$ is the observed frequency and $f_e$ is the expected frequency under the null hypothesis of the Chi-squared test.

association between social and intellectual communities seems hence to exist but it is far from being perfect.

The classification of papers into different intellectual communities permits also to break down the mentions received by each acknowledgees by the community from which they are generated. Table 6 shows the mention decomposition for the acknowledgees with more than 20 mentions.

| **Rank** | **Acknowledgee** | **Intellectual communities** | | | | | | | **Tot** | **N distinct communities** |
|---|---|---|---|---|---|---|---|---|---|---|
| | | ***1*** | ***2*** | ***3*** | ***4*** | ***5*** | ***6*** | ***7*** | | |
| 1 | Katz Lawrence F. | 4 | 11 | 6 | 1 | 4 | 23 | 16 | **65** | 7 |
| 2 | Shleifer Andrei | 0 | 8 | 12 | 4 | 12 | 22 | 2 | **60** | 6 |
| 3 | Shapiro Jesse M. | 1 | 4 | 13 | 2 | 2 | 19 | 6 | **47** | 7 |
| 4 | Acemoglu Daron | 4 | 11 | 0 | 8 | 2 | 14 | 3 | **42** | 6 |
| 5 | Chetty Raj | 4 | 3 | 7 | 0 | 2 | 10 | 13 | **39** | 6 |
| 6 | DellaVigna Stefano | 1 | 1 | 18 | 3 | 0 | 9 | 2 | **34** | 6 |
| 7 | Heckman James J. | 12 | 0 | 4 | 0 | 0 | 6 | 11 | **33** | 4 |
| 8 | Samuelson Larry | 3 | 1 | 5 | 17 | 0 | 2 | 0 | **28** | 5 |
| 9 | Kline Patrick | 9 | 4 | 2 | 0 | 3 | 0 | 9 | **27** | 5 |
| 9 | Finkelstein Amy | 2 | 7 | 5 | 0 | 0 | 7 | 6 | **27** | 5 |
| 9 | Card David | 4 | 3 | 3 | 0 | 2 | 4 | 11 | **27** | 6 |
| 10 | Horner Johannes | 1 | 2 | 1 | 20 | 0 | 1 | 0 | **25** | 5 |
| 10 | Glaeser Edward L. | 4 | 4 | 6 | 1 | 4 | 3 | 3 | **25** | 7 |
| 10 | Gentzkow Matthew | 0 | 5 | 6 | 3 | 2 | 5 | 4 | **25** | 6 |
| 11 | Gabaix Xavier | 0 | 3 | 5 | 4 | 11 | 0 | 1 | **24** | 5 |
| 12 | Bonhomme Stephane | 12 | 3 | 2 | 0 | 2 | 1 | 3 | **23** | 6 |
| 13 | Notowidigdo Matthew J. | 3 | 3 | 3 | 1 | 2 | 3 | 7 | **22** | 7 |
| 13 | Fudenberg Drew | 0 | 1 | 7 | 13 | 1 | 0 | 0 | **22** | 4 |
| 13 | Bloom Nicholas | 0 | 10 | 0 | 1 | 8 | 2 | 1 | **22** | 5 |
| 14 | Laibson David | 1 | 0 | 14 | 0 | 0 | 4 | 2 | **21** | 4 |
| 14 | Banerjee Abhijit | 0 | 2 | 2 | 4 | 0 | 13 | 0 | **21** | 4 |

**Table 5: Mention decomposition for the acknowledgees with more than 20 mentions.** *The total number of papers that mention each acknowledgee is broken down by the intellectual community they are classified into. The community is attributed based on bibliographic coupling similarity between papers. For instance, 4 papers mentioning Lawrence Katz are classified in intellectual community number 1.*

Interestingly, even if some of the top acknowledgees seem to be particularly associated with some community, most of them receives their mentions from 6 or 7 communities.

In fact, in the overall population of acknowledgees, the number of mentions results to be positively correlated with the number of distinct mentioning communities ($R^2 = 0.61$), i.e., the more mentions an acknowledgee collects, the more numerous are the communities mentioning them.

According to the strategic account, the dispersion of the mentions of the top acknowledgees in various intellectual communities shows that mentions are given independently of the intellectual specialization of the acknowledgees. The defender of the normative account, on the other hand, may reply that each acknowledgee receives most of their mentions from some specific communities, and not randomly from any community.

# 7. Discussion

The normative and the strategic account offer contrasting interpretations of the function of the acknowledgments in the primary communication system of science. According to the normative account, the acknowledgments serve to repay debts towards informal and sometimes even formal collaborators. Therefore, they are a reliable source for studying informal collaboration in science. According to the strategic account, on the other hand, the acknowledgments serve to increase the perceived quality of papers by associating the authors to important names in the disciplines. Therefore, they do not automatically attest collaboration. To assess which of these two interpretations is more plausible, six characteristics of the aggregated acknowledgment behavior of economists are investigated.

The stylistic analysis of texts reveals that most of the acknowledgments contain words denoting peer interactive communication (80.7%, a higher proportion than the 60% found by Costas and Leuween but below the 92.2% reported by Alvarez and Caregnato, see Table 3). The high occurrences of words such as “comment”, “seminar”, “participant”, and “assistance” (Table 2), as well as the significant number of texts mentioning seminars and conferences (Table 3) highlight the role that presentations and exchange of ideas play in the acknowledgments in economics, a result that is consistent with previous studies (Alvarez & Caregnato, 2020; Brown, 2005; Rose & Georg, 2021).

These findings seem to support the normative account, in so far economists privilege a thanks-related vocabulary. However, according to the strategic account, the style would not show that acknowledgments are not signals, rather that signals are to be sent following a precise academic etiquette. Accordingly, the mere occurrence of words denoting peer interactive communication does not guarantee that such a communication happened or was relevant. It only says that strategic mention of important scholars should fit the academic conventions. By the same token, conferences and seminars may be mentioned to signal the competence of the authors more than to recognize the contribution of participants and audiences.[11]

From this point of view, the strategic account is supported by the identity of the most mentioned acknowledgees. The top ranks of the list include several economists with prestigious and influential positions at the top American universities and that are in the editorial boards of the T5 as well (Table 4). This result is easily explained within the strategic framework, as the signaling value of these influential researchers is high. By associating their own work with these figures, authors increase their own symbolic capital and the chances of being read and cited.

The same reasoning can be applied to explain the skewed distribution of mentions (Figure 1), which is in line with previous studies of acknowledgments (Alvarez & Caregnato, 2021; Cronin et al., 2003; Giles & Councill, 2004; McCain, 2018). If economists trade on the symbolic capital of the acknowledgees, they will tend to mention those already mentioned in a dynamic of *cumulative advantage* (Price, 1976) analogous to the Matthew effect (Merton, 1988). By contrast, according to the normative account, the concentration pattern should be interpreted as reflecting a great inequality in the propensity to collaborate among acknowledgees or the fact that authors search frequently help and advice in a rather narrow fraction of the population. Even if differences in intellectual specialization may account for some variation in the mentions, as certain economic specialties are likely to attract more informal collaboration, the identity of the most mentioned acknowledgees does not reveal any specific intellectual preference.

At the same time, however, less than 1 acknowledgee out of 5 in the average acknowledgement is highly mentioned, whereas almost half of the acknowledgees are 1-mentioned researchers. This result fit badly with the strategic account, because a high proportion of the acknowledgees of the average paper seem not to be influential. Note, however, that a scholar may be very influent in a local setting and, hence, induce few but nonetheless strategic mentions. In other words, it cannot be concluded from the fact that an acknowledgee receives few mentions that these mentions lack strategic motivations. Signals are not necessarily directed to the big community; they may be sent to small audiences.

The application of the symmetric-acyclic decomposition to the acknowledgment network individuated a hierarchical structure of clusters of scholars, and a prevalence of not-reciprocated acknowledgements pointing toward the highest level of the hierarchy (Figure 2 and 3). In particular, a relatively small central cluster of peers emerged as the central structure of the network (Figure 4). It contains the most mentioned scholars of the whole network and many of them are also editors of the T5. This configuration appears at odds with the normative account, which predicts a prevalence of reciprocating acknowledgements and a "flat" hierarchy of scholars.

[11] Similarly, Aagaard and colleagues (2021) note that funding may be over-represented «to boost apparent outcomes of grants and/or author reputations (by over-emphasizing or even spuriously naming prestigious funders) including when little or no relationship exists between acknowledged funding and the actual published research» (p. 9).

Lastly, the correlation between the acknowledgment coupling and the bibliographic coupling matrices and the comparison of their community structure (Table 5 and Figure 5) seem to show that both intellectual and social factors intervene in the choice of the acknowledgees. In fact, if the acknowledgees would be chosen only in function of their intellectual contribution, the social similarity based on the shared acknowledgees between papers would be highly correlated with the intellectual similarity based on the shared cited references between papers, because both the acknowledgees and the cited references would reflect the scientific topic of the paper. On the other hand, the view that only strategic considerations guide the choice of the acknowledgees scarcely fit with the data, as it would imply that the two networks would have completely unrelated structures.

# 8. Conclusion

In this paper, we addressed the value of acknowledgments as a source of information about scientific collaboration. We analyzed the acknowledgments of 1218 articles published in the so-called "top-five journals" of economics. Since our analyses were performed only on the discipline of economics and its elite journals, further research is needed to better understand whether and how our findings can be generalized to other areas of economics or other academic disciplines. The acknowledgments behavior may vary significantly among fields (see Paul-Hus, Díaz-Faes, et al., 2017). Bearing these caveats in mind, our results permit nonetheless to discuss the normative and the strategic account of the acknowledgements. In particular, the mixed results obtained suggest that neither the normative nor the strategic account offer complete explanations of the observed properties of the acknowledgments. Some characteristics better fit with one or the other framework.

In the light of these results, we propose to stop considering the normative and the strategic account as *mutually excluding alternatives*. They should be conceived as *partial accounts* of the various motivations behind the acknowledging behavior of researchers. Acknowledgements should be intended as the result of a *process*, analogously to what Cronin proposed some time ago for citations (Cronin, 1984). As human products, they are the outcome of different, sometimes overlapping or even contradictory, motivations. The same person can be acknowledged because of an intellectual debt and, at the same time, to signal to the editor that the author is acquainted with an important figure in the field. Or, in the same acknowledgments, the contribution of some influent person may be overemphasized by, at the same time, recognizing the help of fellow researchers. Based on our data, we can say that the mentions of highly mentioned researchers are most likely explained by a cumulative advantage mechanism based on influence rather than by the concentration of collaborative attitude in a few individuals. At the same time, it is implausible that authors mention other researchers *only because of* strategic considerations.

Therefore, the normative and the strategic account should be intended as *valid but partial explanations* of acknowledgments data. A moderate acceptance of both accounts should then be the right attitude to adopt. This position has several advantages. On the one hand, the moderate acceptance of the strategic account reminds the analysist that acknowledgments do not reflect *only* informal collaboration. In the contemporary age of performance-based research evaluation and ubiquitous rankings, such a reminder may help not to embrace too enthusiastically the project of acknowledgments-based metrics of researchers. Once provided with a clear normative value as indicators, it is easy to imagine how acknowledgments could be strategically adapted to score better on metrics, as it has partially happened with citations (Biagioli & Lippman, 2020). If the opportunistic use of references to boost citation metrics can be potentially controlled by referees during the review process, as perfunctory references can be spotted out, a control on strategic acknowledgements would be in fact impossible without a detailed knowledge of the social network of the authors of the paper. Ironically, providing the acknowledgments with a normative value would render the strategic account the *full* explanation of acknowledgment behavior. Moreover, a survey revealed that researchers seem not to approve the use of acknowledgments counting as a further indicator of academic impact (Alvarez & Caregnato, 2020). On the other hand, the moderate acceptance of the normative account can justify the interest of the acknowledgments as objects of study for *descriptive* quantitative studies of science. In those fields, such as economics, where co-authorship is less diffuse, they can offer insightful data to trace informal collaboration that would be difficult to obtain with other methods.

From a methodological and theoretical point of view, the main contribution of this paper is to highlight the complex and multi-dimensional nature of acknowledgment data. They can neither be taken as fully transparent narratives of scientific collaborative relations nor be discarded as mere instrumental devices completely disconnected from collaboration practices.

More generally, we think that our results point out the *stratified nature* of scientific documents and their features. Elements such as the by-line, the acknowledgments or the cited references stand at the crossroad of multiple social and epistemic dynamics which often render univocal interpretations of them reductive or one-sided. The relationship between these textual traces and high-level socio-epistemic phenomena such as scientific collaboration or scientific impact is not so linear or transparent as policymakers involved in science management often desire. However, we think that highlighting this complexity is among the key contributions of quantitative studies of science and should not be easily sacrificed.

Our results suggest several directions for future research on the multi-dimensional nature of the acknowledgments. A possible development is to extend the present analyses to other areas of economics (e.g., journals outside the Top 5 or non-mainstream journals) and other academic fields as a well to better understand whether our results can be generalized. Moreover, the acknowledgments data contained in other citation databases may be compared with those in Web of Science. In addition, the relation between citation-based networks and acknowledgements-based networks is worth further exploration. In the present paper, we compared the bibliographic coupling network with the acknowledgment coupling network, finding a medium association between the structure of the two. However, further networks can be extracted from the acknowledgements and compared with their citation-based counterparts: for instance, the network of co-mentioned acknowledgees may be compared with the network of co-cited authors to investigate whether social and intellectual proximity are correlated. Lastly, the centralities of scholars in the various networks (acknowledgements, citation, co-authorship, etc.) may be investigated to shed light on their socio-intellectual trajectories along the different dimensions of the research activity.

# Appendix

The triadic census of the acknowledgment network produced with *Pajek* is reported below.

| **Type** | | | **Number of triads (ni)** | **Expected (ei)** | **(ni – ei) / ei** | **Model** |
|---|---|---|---|---|---|---|
| 3 | – | 102 | 3,215,762 | 69410.0. | 45.33 | Balance |
| 16 | – | 300 | 10 | 0.00 | 11,049,767,037.59 | Balance |
| 1 | – | 003 | 116,988,297,532 | 116,984,419,091.93 | 0.00 | Clusterability |
| 4 | – | 021D | 477,196 | 69410.03 | 5.88 | Ranked Clusters |
| 5 | – | 021U | 186,699 | 69410.03 | 1.69 | Ranked Clusters |
| 9 | – | 030T | 12,165 | 61.74 | 196.05 | Ranked Clusters |
| 12 | – | 120D | 749 | 0.01 | 54,560.54 | Ranked Clusters |
| 13 | – | 120U | 885 | 0.01 | 64,467.58 | Ranked Clusters |
| 2 | – | 012 | 304,448,721 | 312,151,732.07 | -0.02 | Transitivity |
| 14 | – | 120C | 403 | 0.03 | 14,677.44 | Hierarchical Clusters |
| 15 | – | 210 | 200 | 0.00 | 16380166.67 | Hierarchical Clusters |
| 6 | – | 021C | 241,140 | 138,820.06 | 0.74 | Forbidden |
| 7 | – | 111D | 13,976 | 61.74 | 225.38 | Forbidden |
| 8 | – | 111U | 21,843 | 61.74 | 352.81 | Forbidden |
| 10 | – | 030C | 221 | 20.58 | 9.74 | Forbidden |
| 11 | – | 201 | 578 | 0.01 | 42,103.90 | Forbidden |
| Transitive | | | 13,809 | 61.76 | | |
| Intransitive | | | 278,361 | 138,964.15 | | |